# Symbolic Representation for Analog Realization of A Family of Fractional Order Controller Structures via Continued Fraction Expansion

Anindya Pakhira[a], Saptarshi Das[b], Indranil Pan[c], and Shantanu Das[d]

a) Department of Instrumentation and Electronics Engineering, Jadavpur University, Salt Lake Campus, LB-8, Sector 3, Kolkata-700098, India.
b) School of Electronics and Computer Science, University of Southampton, Southampton SO17 1BJ, United Kingdom.
c) Department of Earth Science and Engineering, Imperial College London, Exhibition Road, London SW7 2AZ, United Kingdom.
d) Reactor Control Division, Bhabha Atomic Research Centre, Mumbai-400085, India.

**Authors' Emails:**

anindya.pakhira@gmail.com (A. Pakhira)

s.das@soton.ac.uk, saptarshi@pe.jusl.ac.in (S. Das*)

i.pan11@imperial.ac.uk, indranil.jj@student.iitd.ac.in (I. Pan)

shantanu@ barc.gov.in (Sh. Das)

**Phone no.** +44-7448572598

**Abstract:**

This paper uses the Continued Fraction Expansion (CFE) method for analog realization of fractional order differ-integrator and few special classes of fractional order (FO) controllers *viz.* Fractional Order Proportional-Integral-Derivative (FOPID) controller, FO[PD] controller and FO lead-lag compensator. Contemporary researchers have given several formulations for rational approximation of fractional order elements. However, approximation of the controllers studied in this paper, due to having fractional power of a rational transfer function, is not available in analog domain; although its digital realization already exists. This motivates us for applying CFE based analog realization technique for complicated FO controller structures to get equivalent rational transfer functions in terms of the controller tuning parameters. The symbolic expressions for rationalized transfer function in terms of the controller tuning parameters are especially important as ready references, without the need of running CFE algorithm every time and also helps in the synthesis of analog circuits for such FO controllers.

*Keywords*: Analog realization; continued fraction expansion (CFE); domino-ladder; fractional order controller; FO[PD] controller; FO lead-lag compensator; symbolic realization

# 1. Introduction

Fractional Calculus is a considerably old field of Mathematics which has witnessed growing interest in recent times from the Science and Engineering community. It has been established that many real world dynamical systems can be better characterized by fractional order models and are governed by fractional order differential equations [1]. Also, fractional order $PI^\lambda D^\mu$ type controllers have been shown to outperform the conventional PID controllers due to the extra tuning knobs and the consequent better flexibility to meet time or frequency domain design specification [2]. Therefore, recent research results have been focused on analog or digital implementation (realization) of fractional order controllers [3]. It is well known that the building blocks of FO controllers in different structures are the fractional order differ-integrators which are basically infinite dimensional linear filters. Hence their band limited practical realization has always been an issue [3–8]. In band limited implementation of the FO elements, care should be taken about the chosen frequency band so that it encompasses the frequency range of operation for specific applications like process control (few milli-Hertz) to filtering of radio-frequency waves (few kilo-Hertz) etc. This indicates that FO elements need to be approximated with finite dimensional transfer functions in a specified frequency-band of interest. Therefore methods are being researched in order to develop such approximations of these FO systems which can be realized in hardware while keeping the order of the filters as low as possible. In this paper, analog realization of few FO controllers, having relatively complex structure is reported.

FO systems can be realized on hardware using two basic approaches *viz.* digital realization and analog realization. In case of digital realization, an equivalent Finite Impulse Response (FIR) or Infinite Impulse Response (IIR) filter can be designed which mimics the response of the FO system [9], using expansion of various generating functions, like Tustin, Euler, Al-Alaoui etc. [10], [11]. Due to the versatility and reliability of modern digital signal processors, digital realization has come up as a viable FO hardware realization method. Analog realization entails designing an equivalent analog electronic circuit, consisting of passive/active components, which mimics the response of the FO system. Digital realization of FO operators can be done in direct or indirect way. In direct realization, we investigate methods that directly lead to a digital realization of a FO transfer function. While in indirect realization, an analog version of the FO transfer function is realized by frequency domain fitting of ideal response of the FO operator which is subsequently discretized by an appropriate $s \leftrightarrow z$ transform. In the popular methods of digital realization, the FO operators are approximated by suitable generating functions, followed by expansion by a Power Series Expansion (PSE) or CFE. PSE leads to FIR-like structures, which are comparatively less efficient due to higher order of resulting filters, while CFE leads to more efficient IIR-like structures. In Vinagre *et al.* [12], digital realization was achieved by expansion of the Tustin operator by Muir recursion and then CFE, which exhibits large errors in high-frequency range. This was later improved in [13] by the use of Al-Alaoui operator which is a combination of Euler and Tustin operators. In [14], CFE based hybrid digital FO integrator and differentiator have been devised which use a combination of the Simpson and Trapezoidal digital integrators as the generating function. Petras in [15] discussed about the hardware implementation of these various types of discrete FO controllers on programmable logic controllers and microcontrollers.

Even though digital methods are inherently more versatile, analog realization is preferable in a few cases like the following:

- Use of digital controllers is almost impossible for extremely fast processes like vibrations. Here analog controllers have to be used.

- Analog realization affords better prospects of analysis *viz.* pole/zero, bandwidth, stability analysis etc.

- Analog realization is required for indirect discretization to obtain digital controllers [11] where a higher order continuous time transfer function is first designed to maintain the desired constant phase within a chosen frequency band. Then it is discretized with a specified sampling time, depending on the nature of application.

Quite a few analog realization methods have been devised which approximate the response of FO operators of the form $s^{\pm\lambda}, \lambda \in \Re$ *viz.* Carlson's method, Oustaloup's method, Matsuda method and the Charef's method etc. [3], [4], [16]. However, it has been shown by Das *et al.* [17] that the higher order Carlson's realization method provides satisfactory results for semi-differentiator only. Charef's method [18] is based on magnitude curve fitting and hence, phase curve fitting is poor in this case. For good magnitude fitting with specified error bound in dB, Charef's method results in very higher order approximation which is not preferable [8], [18]. Oustaloup's method generally gives good frequency domain fitting for constant phase fractional order elements within chosen frequency band [19]. Near the low and high frequency boundaries, the Oustaloup's approximation is not so good and its improved version known as modified Oustaloup's approximation method [20] has the capability to overcome these shortcomings. Also, with both the Oustaloup's methods, to maintain large constant phase region, the phase ripples increases even with high order approximations. Thus to achieve satisfactory performance for wider frequency range, order of the obtained analog approximation becomes very high, which is not suitable from analog circuit implementation point of view. Recommended setting for Oustaloup's method for FOPID controller design has been discussed by Bayat [21] and a simulated annealing based optimization framework for parameter selection of modified Oustaloup's method has been shown by Das *et al.* [22]. In fact, very few of the above methods can be reliably applied for complicated fractional order structure containing fractional power of a rational transfer function like FO lead-lag compensator [23], Fractional Order [Proportional Derivative] or FO[PD] and Fractional Order [Proportional Integral Derivative] or FO[PID] controller etc. in analog domain [24], [25]. This is the motivation of applying CFE to produce iterative rational approximations for such irrational transfer functions.

In [8], [11], a CFE based method for the realization of FO operators has been proposed. The technique relies on a suitable generating function which is equivalent to the FO operator, which is then expanded to obtain the required approximation, in rational transfer function form. Using Power Series Expansion (PSE) on the other hand leads to discrete time transfer functions in the form of polynomials. Additionally, it is discussed in [26] that CFE often converges more rapidly than PSE. In [3] the magnitude and phase responses for rational approximations based on CFE and the Oustaloup's method have been compared, and the CFE based approximation has been observed to be closer to the ideal case. In Dorcak *et al.* [6] a CFE based realization of analog FO systems using the FO operator has been studied, along with its hardware implementation. The convenience of the truncated CFE in obtaining coefficients of the domino ladder based realization directly, without any transformation, has been relied upon. Even CFE based digital realization for the $s^{\pm\lambda}$ operator, when compared to the magnitude and phase response of an ideal FO integro-differential operator, is not satisfactory. The frequency band for which the realization is valid is limited, even for higher order realizations. Hence, in [27], a Particle Swarm Optimization (PSO) based framework utilizing CFE is presented, where the parameters of a generalized truncated CFE are made to fit the ideal response, by minimizing the difference in magnitude and phase response of the ideal and realized transfer functions. Another approach, using genetic algorithm for optimizing the weights of the averaging of different generating functions, in order to balance high and low frequency accuracies, is proposed in [10]. In [28], cascaded two-port networks, each consisting of an operational amplifier, resistor and capacitor, have been used to realize FO differentiator and integrator with constant phase response over a specific frequency range. Besides these methods based on CFE, PSE, etc. and the resulting complex electrical networks, a relatively new way has been realized for FO memristive circuits in [5], [29]. The circuit used is a simple operational amplifier in inverting amplifier configuration. The use of these FO elements to implement an FOPID controller has also been proposed.

A limitation of all the above mentioned analog realization techniques is that they are valid and has been reliably used only for the nominal $s^{\pm\lambda}$ operator. This allows implementation of the FOPD, FOPI and FOPID controllers. However, there are certain class of FO controller structures which cannot be realized using these approximation, $viz.$ FO[PD], FO[PI], FO[PID] and FO lead-lag compensator etc. [23–25]. In [25], an FO[PD] controller has been realized in the digital domain, using the impulse response invariant discretization method. This is a time-domain approach, where the impulse response of the ideal FO[PD] controller is sampled at a certain sampling frequency and a digital filter is realized which mimics the time domain response. However, analog realization of the same controller has not yet been studied except using Carslon's method in [17]. In the present paper, analog realization of a few special classes of FO controllers have been attempted $viz.$ FO[PD] controller and FO lead-lag compensator in parametric form. Moreover, with increase in the order of the truncated CFE and the resulting rationalized transfer function, accuracy of the realization generally increases but it might affect the practical realizability in hardware due to the large size of the network. Therefore, a trade-off between accuracy and increase in the order of approximation (model complexity) has been investigated and an empirical optimum result has been opted here. Also, the rational approximation of such fractional order controllers with fixed (tuned) parameters leads to higher order transfer functions and the associated coefficients become widely different with the order of realization and the controller tuning parameters (knobs). Therefore, after studying the frequency domain fitting characteristics of such controllers symbolic transfer functions are obtained via CFE which will help in obtaining the hardware realization for different controller parameters. This is particularly important for industrial users as the symbolic transfer functions after realization facilitates calculation of the realized higher order transfer functions using only the controller/compensator gain and orders, without the need of running CFE every time. In the present design, the rationalized higher order transfer functions balances the accuracy of frequency domain fitting as well as less complex realization. The contribution of this article is firstly to study the CFE based analog realization of complicated controller structures like fractional powers of rational transfer functions $\left[G\left(s\right)\right]^{\pm\lambda}$ and then choosing a low order but high accuracy trade-off for the rational approximation. Secondly it also generates symbolic expressions for those controller structures in terms of their tuning parameters. Although the CFE based rational approximation techniques for FO controllers are sometimes complained to have a lower band-width than that with the other approximation techniques [3], still it is quite popular where a narrow band-width operation can be tolerated [11][26].

The rest of the paper is organized as follows. Section 2 describes the basics of CFE based analog realization for the FO controllers. Section 3 describes the simulations and empirical rationalized forms for the three FO controllers. The paper ends with the conclusion as section 4, followed by the references.

## 2. CFE based analog realization for fractional order controllers

### 2.1. Basics of continued fraction expansion

Continued fraction expansion is an iterative way of representing a number or a function in the form given in (1).

$$r = a_0 + \cfrac{b_1}{a_1 + \cfrac{b_2}{a_2 + \cfrac{b_3}{a_3 + \ldots}}} \tag{1}$$

where, $a_i, b_i$ are either real or complex numbers. The expansion may be infinite or truncated to a finite expression. If $b_i = 1$ for all $i$, then the CFE is known as simple CFE. The numbers $a_i$ are known as partial coefficients. If a CFE is truncated to an expression having $k$ partial coefficients it is known as the $k^{th}$ convergent of the continued fraction.

### 2.2. CFE based approximation of FO transfer functions

CFE has been used in the past as a method of approximation of functions. It shows faster convergence as compared to PSE and also has a wider domain of convergence in the complex plane [26], [30] and consequently require less number of coefficients to obtain an acceptable approximation. Therefore, CFE can be used to approximate irrational functions. Hence, FO transfer functions, which are basically a class of irrational functions, can be approximated by using CFE. The CFE based approximation of an irrational transfer function $G(s)$ can be expressed as (2).

$$G(s) \simeq a_0(s) + \cfrac{b_1(s)}{a_1(s) + \cfrac{b_2(s)}{a_2(s) + \cfrac{b_3(s)}{a_3(s) + \cdots}}} \tag{2}$$

where, $a_i(s), b_i(s)$ are functions of Laplace variable 's' or are constants. From CFE, convergent polynomials corresponding to the numerator and denominator of a rational function can be obtained. This function represents the final rational approximation of the irrational transfer function $G(s)$. It is to be noted that the $n^{th}$ convergent of a CFE results in an $(n/2)^{th}$ order approximation. The FOPI, FOPD and FOPID controllers can be realized in hardware by using the approximation for the $s^{\pm\lambda}$ operator as described in [4], [8]. The present work focuses on a method to realize primarily complicated structures like FO[PD] controllers and the FO lead-lag compensator. Approximation of these two controller transfer functions has been attempted by directly obtaining the CFE of the original transfer functions and truncating the CFE to an order which shows acceptable accuracy. The magnitude and phase responses for different orders are obtained and compared with ideal responses. Finally, symbolic transfer functions have been provided in terms of the controller tuning parameters for CFE orders, considering a trade-off between accuracy and complexity.

### *2.3. CFE and the domino ladder circuit*

The concept of domino ladder circuit can be used to realize the irrational transfer functions based on CFE with real-world circuit elements. Consider the finite ladder circuit in Fig. 1, where $Z_{2n-1}(s)$ and $Y_{2n-1}(s)$ are the series impedance and shunt admittances of the circuit. Considering the circuit from left-to-right direction, the resulting impedance is given by (3).

$$Z(s) = Z_1(s) + \cfrac{1}{Y_2(s) + \cfrac{1}{Z_3(s) + \cfrac{1}{Y_4(s) + \cfrac{1}{\ddots \cfrac{}{Y_{2n-2}(s) + \cfrac{1}{Z_{2n-1}(s) + \cfrac{1}{Y_{2n}(s)}}}}}}} \tag{3}$$

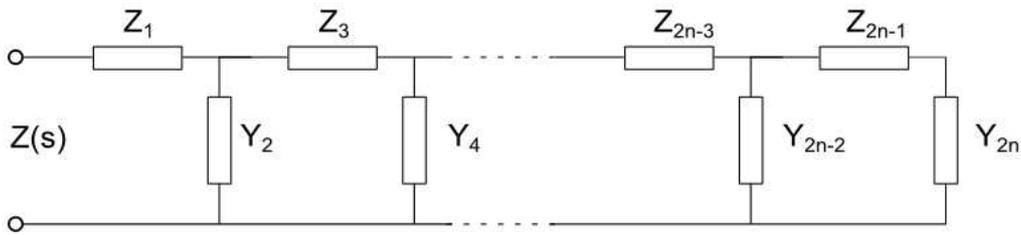

Fig. 1. Ladder circuit implementation for continued fraction expansion.

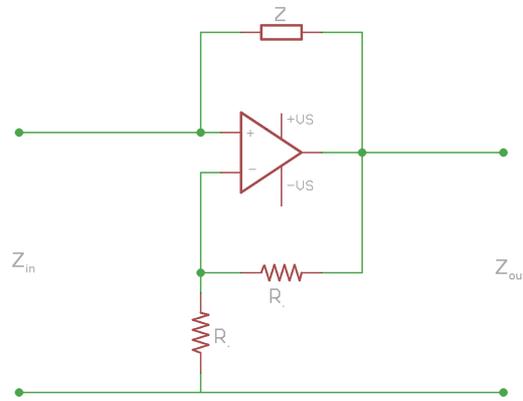

Fig. 2. Typical negative impedance converter circuit.

The impedance of the circuit exactly matches with the form of a CFE. Hence, the physical circuit components, corresponding to a particular CFE based realization of an FO controller can be obtained by computing the CFE of the realized transfer function and comparing with the above equation. Then the comparison with the domino ladder circuit components $Z_{2n-1}(s)$ and $Y_{2n-1}(s)$ will give the required circuit components. The values of impedance/admittance obtained by the given method can come out to be negative. In such cases, negative impedance realization circuits [31] can be used to realize these impedances. In [32], operational amplifier based active circuits known as *negative impedance converters* (NIC) have been detailed, like the one given in Fig. 2.

An NIC is usually a one-port op-amp circuit and works by reversing the voltage or current direction in the port, thereby introducing a 180 degree phase shift between the voltage and current. Fig. 2 shows a negative impedance convertor with current inversion (INIC), the other variety being a negative impedance convertor with voltage inversion (VNIC). Let the input voltage be $v$ and the resulting current through the port be $i$. Due to virtual short, the voltage at the inverting terminal is also $v$, and the output voltage is $2v$, due to the voltage divider with equal resistances $R$. Now the current through the impedance $Z$ can be calculated as $i = (v - 2v)/Z \implies v/i = -Z$. Besides the domino ladder circuit, there are other circuit configurations like the RC transmission line circuit (also known as symmetrical domino ladder circuit) and the RC binary tree circuit, as given in [6].

## 3. Simulation and results

### 3.1. Analog realization of FO differintegrators

The $s^{\pm\lambda}$ operator cannot be directly expanded by CFE. As given in Podlubny *et al.* [4], a generating function for the operator is to be used, which is expanded by CFE. Since the generating function approximates the response for a relatively narrow frequency range, two generating functions should be used, one for the low frequency range and one for the high frequency range, as necessitated by the application. The generating functions for an FO integrator $(1/s^\lambda, \lambda \in \Re_+)$, for low and high frequency ranges are given in (4) and (5) respectively.

$$C_{low} = \left(1 + \frac{1}{s}\right)^\lambda, \text{ when } \omega << 1 \tag{4}$$

$$C_{high} = \frac{1}{(1 + sT)^\lambda}, \text{ when } \omega T >> 1 \tag{5}$$

Symbolic expressions for 4th and 5th order, both for low and high frequency ranges are given in (6)-(7) and (8)-(9) respectively. The transfer function for high frequency range has been computed with $T = 1$ [4]. In the higher order transfer function representation the subscripts denotes the order of realization and the superscript denote the frequency range of fitting (i.e. high or low). Expressions (6)-(9) are particularly important since with the knowledge of fractional order $\lambda$, one can easily get the realized higher order transfer functions for simulation or for the purpose of physical circuit implementation. As expected the 5th order symbolic representations are slightly more complex than that with the 4th order CFE, but gives better frequency domain constant phase fitting result. The obtained symbolic expressions can easily be inverted to obtain equivalent expressions for fractional differentiators as well.

$$Q_4^{low} = \frac{(\lambda^3 - 9\lambda^2 + 26\lambda - 24)\binom{1680s^4 + (840\lambda + 3360)s^3 + (180\lambda^2 + 1260\lambda + 2160)s^2}{+(20\lambda^3 + 180\lambda^2 + 520\lambda + 480)s + \lambda^4 + 10\lambda^3 + 35\lambda^2 + 50\lambda + 24}}{(\lambda - 4)(\lambda^2 - 5\lambda + 6)\binom{(1680s^4 + (-840\lambda + 3360)s^3 + (180\lambda^2 - 1260\lambda + 2160)s^2}{+(-20\lambda^3 + 180\lambda^2 - 520\lambda + 480)s + \lambda^4 - 10\lambda^3 + 35\lambda^2 - 50\lambda + 24}}$$

$$(6)$$

$$Q_4^{high} = \frac{\left(\begin{array}{l}(\lambda^4 - 10\lambda^3 + 35\lambda^2 - 50\lambda + 24)s^4 + (-20\lambda^3 + 180\lambda^2 - 520\lambda + 480)s^3 \\ +(180\lambda^2 - 1260\lambda + 2160)s^2 + (-840\lambda + 3360)s + 1680\end{array}\right)}{\left(\begin{array}{l}(\lambda^4 + 10\lambda^3 + 35\lambda^2 + 50\lambda + 24)s^4 + (20\lambda^3 + 180\lambda^2 + 520\lambda + 480)s^3 \\ +(180\lambda^2 + 1260\lambda + 2160)s^2 + (840\lambda + 3360)s + 1680\end{array}\right)} \tag{7}$$

$$Q_5^{low} = -\frac{\left(\lambda^4 - 14\lambda^3 + 71\lambda^2 - 154\lambda + 120\right)\left(\begin{array}{l}30240s^5 + (15120\lambda + 75600)s^4 + (3360\lambda^2 + 30240\lambda + 67200)s^3 \\ +(420\lambda^3 + 5040\lambda^2 + 19740\lambda + 25200)s^2 + (30\lambda^4 + 420\lambda^3 + 2130\lambda^2 \\ +4620\lambda + 3600)s + \lambda^5 + 15\lambda^4 + 85\lambda^3 + 225\lambda^2 + 274\lambda + 120\end{array}\right)}{\left(\lambda - 5\right)\left(\lambda^3 - 9\lambda^2 + 26\lambda - 24\right)\left(\begin{array}{l}-30240s^5\lambda^5 + (15120\lambda - 75600)s^4 + (-3360\lambda^2 + 30240\lambda - 67200)s^3 \\ +(420\lambda^3 - 5040\lambda^2 + 19740\lambda - 25200)s^2 + (-30\lambda^4 + 420\lambda^3 - 2130\lambda^2 \\ +4620\lambda - 3600)s - 15\lambda^4 + 85\lambda^3 - 225\lambda^2 + 274\lambda - 120\end{array}\right)} \tag{8}$$

$$Q_5^{high} = -\frac{\left(\begin{array}{l}(\lambda^5 - 15\lambda^4 + 85\lambda^3 - 225\lambda^2 + 274\lambda - 120)s^5 + (-30\lambda^4 + 420\lambda^3 - 2130\lambda^2 \\ +4620\lambda - 3600)s^4 + (420\lambda^3 - 5040\lambda^2 + 19740\lambda - 25200)s^3 + (-3360\lambda^2 \\ +30240\lambda - 67200)s^2 + (15120\lambda - 75600)s - 30240\end{array}\right)}{\left(\begin{array}{l}(\lambda^5 + 15\lambda^4 + 85\lambda^3 + 120 + 274\lambda + 225\lambda^2)s^5 + (30\lambda^4 + 420\lambda^3 + 2130\lambda^2 \\ +4620\lambda + 3600)s^4 + (420\lambda^3 + 5040\lambda^2 + 19740\lambda + 25200)s^3 + (3360\lambda^2 \\ +30240\lambda + 67200)s^2 + (15120\lambda + 75600)s + 30240\end{array}\right)} \tag{9}$$

In Fig. 3, CFE based low frequency generating function (4) has been realized for a semi-integrator ($\lambda = 0.5$) with various order of realization. It is clear that even though the magnitude responses of the analog approximations have matched the ideal one within the frequency range of 3 milli-Hz to 1 Hz, the phase suffers from maintaining a constant value of -45 degrees. For the phase fitting characteristics, at least 4[th] and 5[th] order realizations are capable of maintaining the desired -45 degrees of phase compared to other lower order realizations. Further increase in the order of CFE, would drastically make the symbolic expressions more complex but we can observe from Fig. 3-4 that the improvement in maintaining constant phase is not much in the low frequency region (upto 1milli-Hz) even for a 10[th] order realization.

Whereas, using the high frequency generating function (5) for fractional integrators the fitting characteristics are reasonable for 4[th] and 5[th] order with the consideration of accuracy and complexity simultaneously. Here, the magnitude curves mimic the ideal 10 dB/decade magnitude roll-off within the range of 2 Hz-1 kilo-Hz. Also, the maximum achievable frequency within which the phase is constant is 1 kilo-Hz. The fourth order CFE based analog realizations due to the fact of being compact and moderate accuracy are therefore applied for representing fractional integrators with various fractional orders ($\lambda$) as shown in Fig. 5-6. The low and high frequency generating function based simulation results are reported in Fig. 5 and 6 respectively which shows that the symbolic representations are valid for analog realization of other fractional powers also i.e. $\lambda = \{0.1, 0.3, 0.5, 0.7, 0.9\}$. It is also observed from Fig. 5-6 that increase in the order of fractional power $\lambda$, shifts the peak of the phase response or the constant phase region towards low and high frequency regions with the use of low and high frequency generating functions respectively. Although in Fig. 5-6, the respective magnitude curves show a constant gain roll off with a slope of -20×$\lambda$ dB/decade for a significantly wide bandwidth but the phase curves shows relatively narrow bandwidth to maintain a constant phase response.

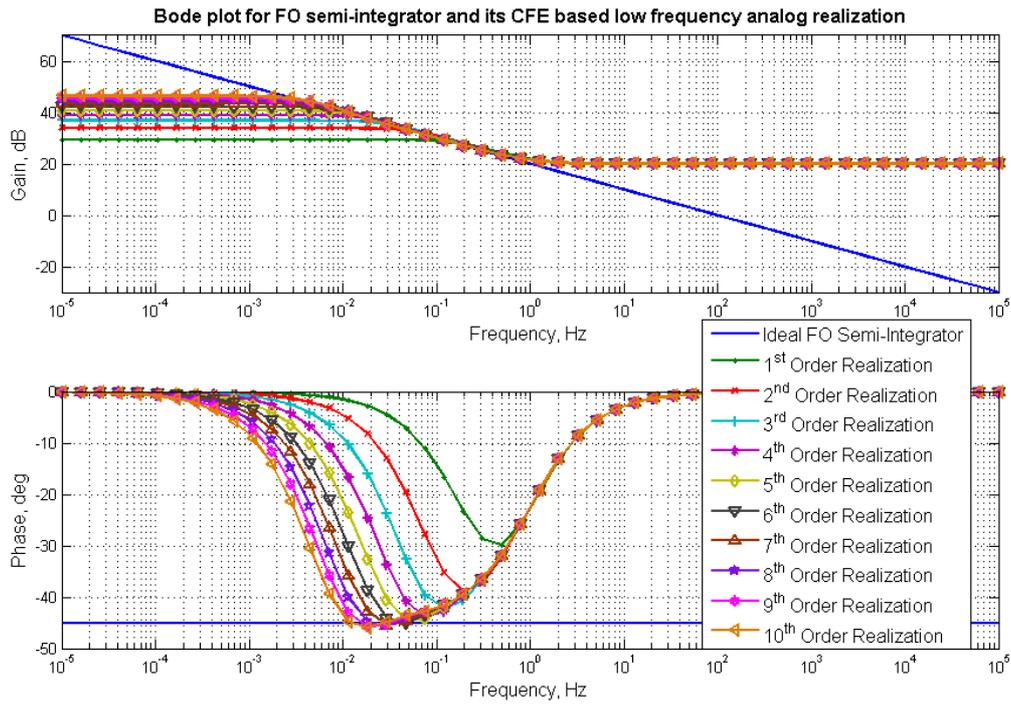

Fig. 3. CFE based low frequency analog approximation of FO semi-integrator.

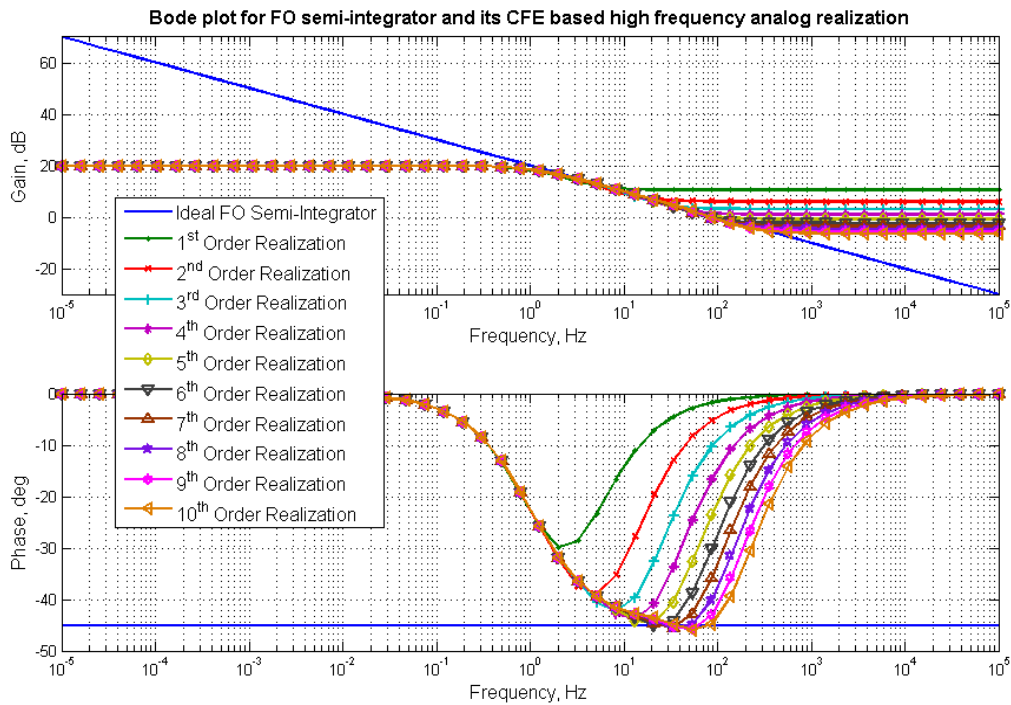

Fig. 4. CFE based high frequency analog approximation of FO semi-integrator.

The CFE-realized 3$^{rd}$ order transfer functions of the $s^{-0.5}$ operator are given as

$$Q_3^{low} = \frac{64s^3 + 112s^2 + 56s + 7}{64s^3 + 80s^2 + 24s + 1} \tag{10}$$

$$Q_3^{high} = \frac{s^3 + 24s^2 + 80s + 64}{7s^3 + 56s^2 + 112s + 64} \tag{11}$$

On performing CFE on $Q_3^{low}$ we have (12).

$$C_3^{low} = 1 + \cfrac{1}{2s + 0.5 + \cfrac{1}{-8s - 4 + \cfrac{1}{2s + 1}}} \tag{12}$$

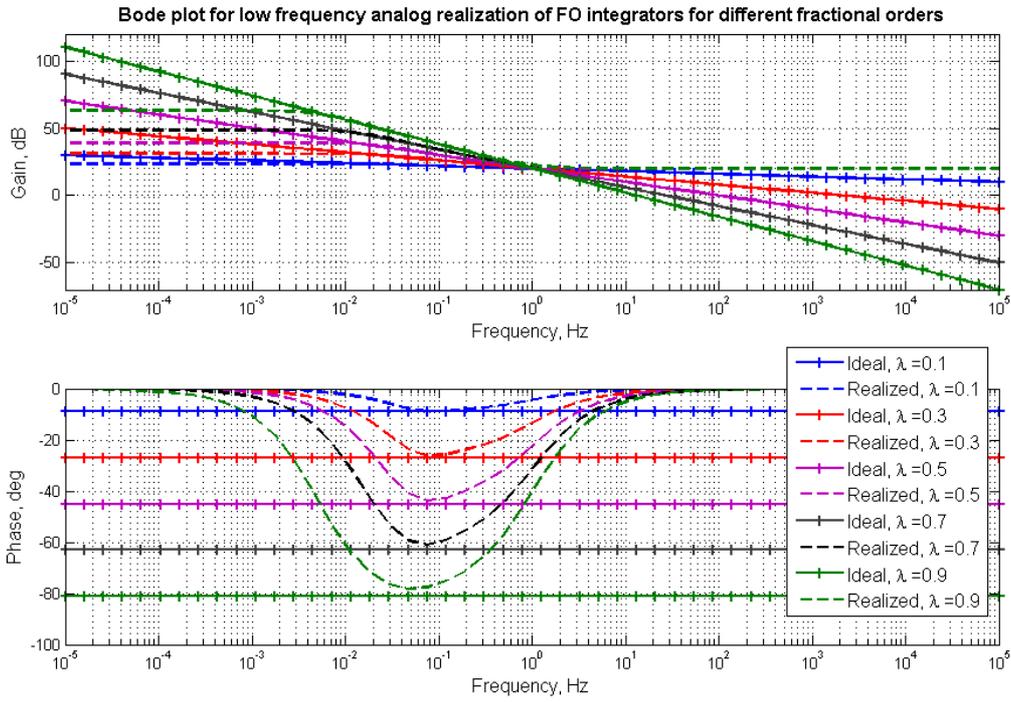

Fig. 5. Fitting characteristics of low frequency approximation for FO integrators at different fractional orders.

Comparing with domino ladder form, the impedance-admittance values are obtained as:

$$\left.\begin{array}{l} Z_1(s) = 1 \\ Y_2(s) = 2s + 0.5 \\ Z_3(s) = -8s - 4 \\ Y_4(s) = 2s + 1 \end{array}\right\}.$$

Whereas, on performing CFE on $Q_3^{high}$ we have

$$C_3^{high} = 0.1429 + \cfrac{1}{0.4375s + 1.75 + \cfrac{1}{-0.6667s - 1.778 + \cfrac{1}{11.81s + 15.75}}} \qquad (13)$$

Comparing with domino ladder form, the impedance and admittance values can be obtained as:

$$\left. \begin{aligned} Z_1(s) &= 0.1429 \\ Y_2(s) &= 0.4375s + 1.75 \\ Z_3(s) &= -0.6667 - 1.778 \\ Y_4(s) &= 11.81s + 15.75 \end{aligned} \right\}$$

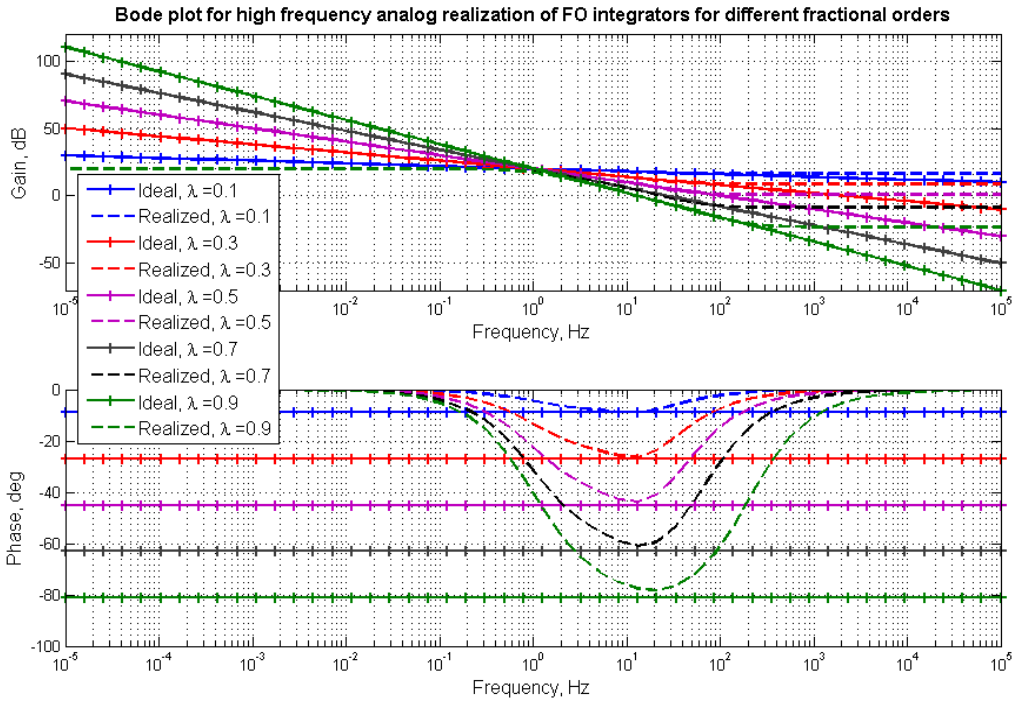

Fig. 6. Fitting characteristics of high frequency approximation for FO integrators at different fractional orders.

It is clear that the obtained impedance and admittances are basically first order lead or lag circuits except the first impedance being simply a gain.

In Fig. 7, we also present a comparison of the frequency responses of FO integrator $1/s^\lambda$ with $\lambda = \{0.1, 0.3, 0.5, 0.7, 0.9\}$ using another three different analog realization methods *viz.* Carlson's, Oustaloup's and modified Oustaloup's method [3]. In all the three cases, we used a third order rational approximation which resulted a 13th, 7th and 9th order transfer function which is still higher than that achieved by the CFE technique. From. Fig. 7, it is evident that using a 3rd order Carlson's method although the magnitude response shows the required gain roll off for the FO integrator within a bandwidth of 0.1-10 Hz, its phase response varies drastically and fail to maintain a constant value. The Oustaloup's method and its modified version in Fig. 7 were reported with a three recursive steps and within a bandwidth of $10^{-3}$-$10^3$ Hz. It is evident that the modified Oustaloup's method has a better constant phase response than that with the Oustaloup's method, especially near the boundaries but both of them suffers from high order of the rationalized transfer function. More details on the relative accuracies of these three techniques could be found in [17], [22].

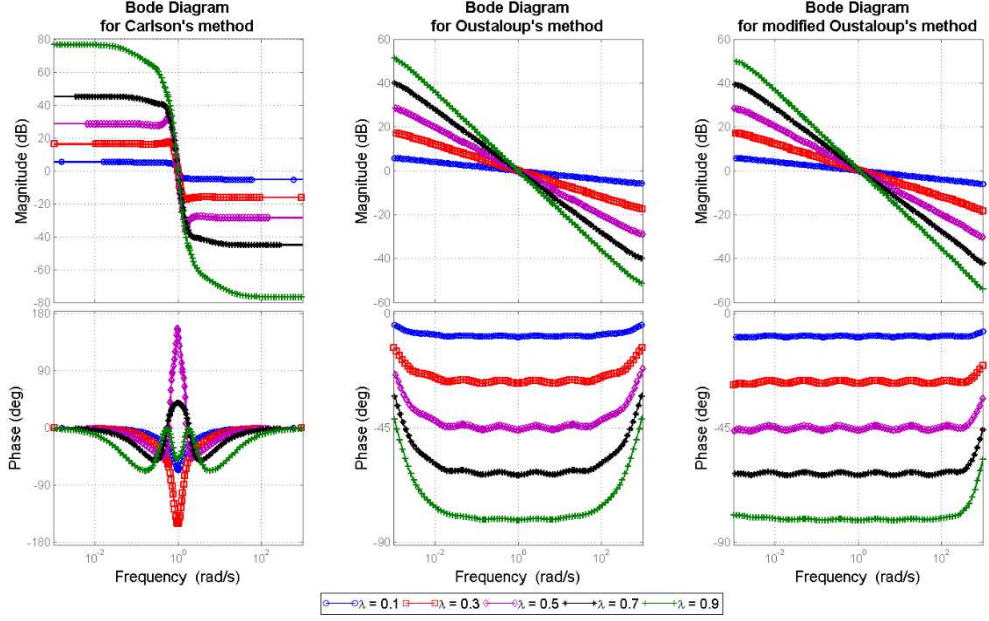

Fig. 7. Comparison of the frequency response of FO integrators using the Carlson's, Oustaloup's and modified Oustaloup's method.

### 3.2. Analog realization of FOPID controller

The transfer function of a generic fractional order proportional integral derivative (FOPID) controller [33] is given as

$$G_{FOPID} = \left( K_p + \frac{K_i}{s^\lambda} + K_d s^\mu \right), \{\lambda, \mu\} \in (0, 2), \{K_p, K_i, K_d\} \in \Re_+ \tag{14}$$

The FOPI controller [34] and FOPD controllers [35] are sub-class of the FOPID controller (14). This can be realized by using the CFE-based realization of the $s^{\pm\lambda}$ operator, obtained via the generating functions for the low- and high-frequency ranges, as detailed in the previous subsection. The 4$^{th}$ order symbolic expressions for the generic FOPID controller are given below with low and high frequency generating functions respectively for the fractional differ-integrators.

$$G_4^{low} = K_p + \frac{K_i\left(\lambda^3 - 9\lambda^2 + 26\lambda - 24\right)\binom{1680s^4 + (840\lambda + 3360)s^3 + (180\lambda^2 + 1260\lambda + 2160)s^2}{+(20\lambda^3 + 180\lambda^2 + 520\lambda + 480)s + \lambda^4 + 10\lambda^3 + 35\lambda^2 + 50\lambda + 24}}{(\lambda-4)\left(\lambda^2 - 5\lambda + 6\right)\binom{(1680s^4 + (-840\lambda + 3360)s^3 + (180\lambda^2 - 1260\lambda + 2160)s^2}{+(-20\lambda^3 + 180\lambda^2 - 520\lambda + 480)s + \lambda^4 - 10\lambda^3 + 35\lambda^2 - 50\lambda + 24)}}$$

$$+ \frac{K_d\left(\mu - 4\right)\left(\mu^2 - 5\mu + 6\right)\binom{(1680s^4 + (-840\mu + 3360)s^3 + (180\mu^2 - 1260\mu + 2160)s^2}{+(-20\mu^3 + 180\mu^2 - 520\mu + 480)s + \mu^4 - 10\mu^3 + 35\mu^2 - 50\mu + 24)}}{\left(\mu^3 - 9\mu^2 + 26\mu - 24\right)\binom{1680s^4 + (840\mu + 3360)s^3 + (180\mu^2 + 1260\mu + 2160)s^2}{+(20\mu^3 + 180\mu^2 + 520\mu + 480)s + \mu^4 + 10\mu^3 + 35\mu^2 + 50\mu + 24}} \tag{15}$$

$$G_4^{high} = K_p + \frac{K_i \begin{pmatrix} (\lambda^4 - 10\lambda^3 + 35\lambda^2 - 50\lambda + 24)s^4 + (-20\lambda^3 + 180\lambda^2 - 520\lambda + 480)s^3 \\ + (180\lambda^2 - 1260\lambda + 2160)s^2 + (-840\lambda + 3360)s + 1680 \end{pmatrix}}{\begin{pmatrix} (\lambda^4 + 10\lambda^3 + 35\lambda^2 + 50\lambda + 24)s^4 + (20\lambda^3 + 180\lambda^2 + 520\lambda + 480)s^3 \\ + (180\lambda^2 + 1260\lambda + 2160)s^2 + (840\lambda + 3360)s + 1680 \end{pmatrix}}$$
$$+ \frac{K_d \begin{pmatrix} (\mu^4 + 10\mu^3 + 35\mu^2 + 50\mu + 24)s^4 + (20\mu^3 + 180\mu^2 + 520\mu + 480)s^3 \\ + (180\mu^2 + 1260\mu + 2160)s^2 + (840\mu + 3360)s + 1680 \end{pmatrix}}{\begin{pmatrix} (\mu^4 - 10\mu^3 + 35\mu^2 - 50\mu + 24)s^4 + (-20\mu^3 + 180\mu^2 - 520\mu + 480)s^3 \\ + (180\mu^2 - 1260\mu + 2160)s^2 + (-840\mu + 3360)s + 1680 \end{pmatrix}} \tag{16}$$

### 3.3. Analog realization of FO[PD] controller

The transfer function of a generic fractional order proportional derivative controller or FO[PD] is given in (17) which is different to that studied in [35]. This typical structure has fractional power of rational transfer function and this typical controller has already been realized by Luo and Chen [25] using impulse response invariant discretization method, which is a digital realization method. Here the same has been achieved in the analog domain.

$$G_{FO[PD]} = \left[ K_p + K_d s \right]^{\mu} = K_p \left( 1 + \frac{K_d s}{K_p} \right)^{\mu}, \quad \mu \in (0,2), \left\{ K_p, K_i \right\} \in \Re_+ \tag{17}$$

Symbolic expressions for 4th and 5th order realizations of an FO[PD] controller are given in (18) and (19) respectively. These expressions are helpful as ready references to obtain rational approximation for such a complicated controller structure.

$$G_4 = \frac{K_p^{\mu} \begin{pmatrix} \left( \mu^4 + 10\mu^3 + 35\mu^2 + 50\mu + 24 \right) K_d^4 s^4 + \left( 20\mu^3 + 180\mu^2 + 520\mu + 480 \right) K_d^3 K_p s^3 \\ + \left( 180\mu^2 + 1260\mu + 2160 \right) K_d^2 K_p^2 s^2 + \left( 840\mu + 3360 \right) K_d K_p^3 s + 1680 K_p^4 \end{pmatrix}}{\begin{pmatrix} \left( \mu^4 - 10\mu^3 + 35\mu^2 - 50\mu + 24 \right) K_d^4 s^4 - \left( 20\mu^3 - 180\mu^2 + 520\mu - 480 \right) K_d^3 K_p s^3 \\ + \left( 180\mu^2 - 1260\mu + 2160 \right) K_d^2 K_p^2 s^2 - \left( 840\mu - 3360 \right) K_d K_p^3 s + 1680 K_p^4 \end{pmatrix}} \tag{18}$$

$$G_5 = \frac{K_p^{\mu} \begin{pmatrix} \left( \mu^5 + 15\mu^4 + 85\mu^3 + 225\mu^2 + 274\mu + 120 \right) K_d^5 s^5 \\ + \left( 30\mu^4 + 420\mu^3 + 2130\mu^2 + 4620\mu + 3600 \right) K_d^4 K_p s^4 \\ + \left( 420\mu^3 + 5040\mu^2 + 19740\mu + 25200 \right) K_d^3 K_p^2 s^3 \\ + \left( 3360\mu^2 + 30240\mu + 67200 \right) K_d^2 K_p^3 s^2 \\ + \left( 15120\mu + 75600 \right) K_d K_p^4 s + 30240 K_p^5 \end{pmatrix}}{\begin{pmatrix} \left( \mu^5 - 15\mu^4 + 85\mu^3 - 225\mu^2 + 274\mu - 120 \right) K_d^5 s^5 \\ - \left( 30\mu^4 - 420\mu^3 + 2130\mu^2 - 4620\mu + 3600 \right) K_d^4 K_p s^4 \\ + \left( 420\mu^3 - 5040\mu^2 + 19740\mu - 25200 \right) K_d^3 K_p^2 s^3 \\ - \left( 3360\mu^2 - 30240\mu + 67200 \right) K_d^2 K_p^3 s^2 \\ + \left( 15120\mu - 75600 \right) K_d K_p^4 s - 30240 K_p^5 \end{pmatrix}} \tag{19}$$

Fig. 8 shows that up to fifth order of realization, improvement in the accuracy of the frequency domain fitting characteristics is clear. But with further higher orders, the resulting models become more complicated and improvement in accuracy is not significant. Also, for simulation of Fig. 8-9, the nominal FO part of the controller has been taken as $K_p = 6.3092, K_d = 0.9435, \mu = 1.205$, as in [25], with the FO part realized using CFE and the integer order part evaluated separately. From both Figs. 8 and 9, it is clear that the frequency domain fitting is good at low frequencies and up to the '$S$'-shaped region of the magnitude curves. For higher frequencies, these do not give a good fit. Thus using such controllers might be useful in industries like chemical process control, which operates at relatively low frequencies. Usage in applications like high frequency power electronic components and electrical drives are not advisable.

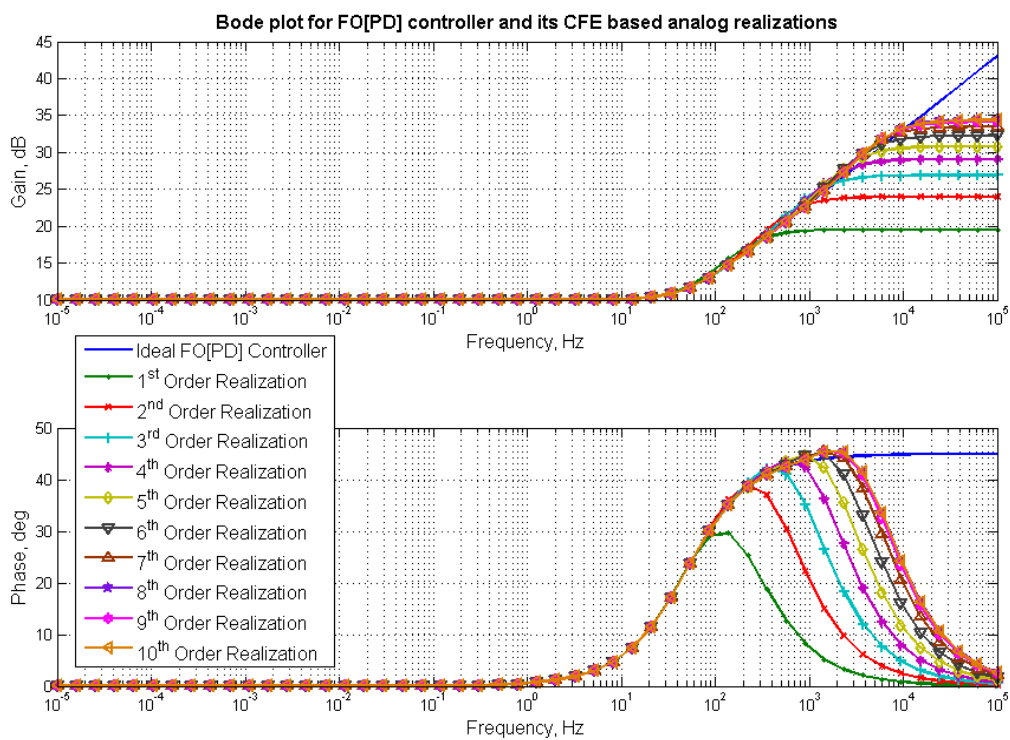

Fig. 8. CFE based analog approximation of FO[PD] controller.

The integral part $(K_p + K_d s)^1$ can be realized by a general active circuit for PD controllers. The impedances and admittances for a 3$^{rd}$ order realization of the fractional order part of the controller from the domino ladder form are arrived at, as shown in (20). The FO part $(K_p + K_d s)^{0.205}$ is realized using CFE. Convergent 3$^{rd}$ order polynomial obtained using CFE:

$$Q(s) = \frac{(2.92s^3 + 30.82s^2 + 74.09s + 49.00)}{(s^3 + 15.99s^2 + 47.23s + 35.82)} \tag{20}$$

This on expanding again by CFE will lead to the required domino ladder form:

$$z_{dom}(s) = 2.92 + \cfrac{1}{-0.06s - 0.75 + \cfrac{1}{3.61s + 9.51 + \cfrac{1}{-1.84s - 2.56}}} \tag{21}$$

Comparing with the domino ladder form, we have

$$\left.\begin{array}{l} Z_1(s) = 2.9208 \\ Y_2(s) = -0.0629s - 0.7536 \\ Z_3(s) = 3.6189s + 9.5151 \\ Y_4(s) = -1.8466s - 2.5622 \end{array}\right\}$$

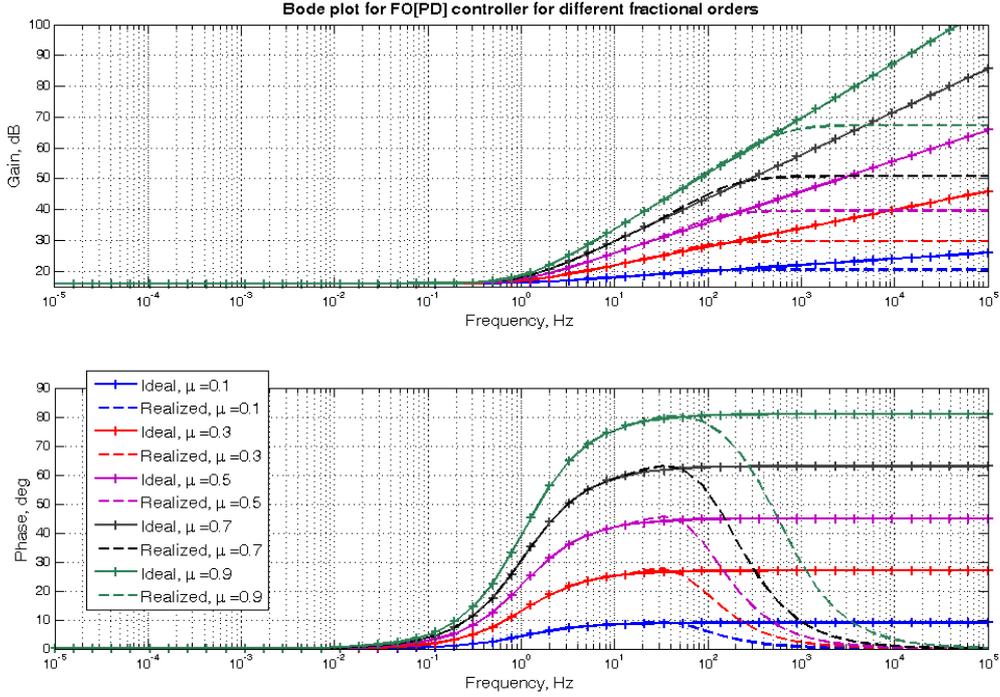

Fig. 9. Frequency domain fitting at different $\mu$ for FO[PD] controller.

### 3.4. Analog realization FO lead-lag compensator

The transfer function of a generic FO lead-lag compensator is given by (22). For numerical simulation, the corresponding parameters of the FO lead-lag compensator have been taken as $K = 141.4214, \lambda = 0.6404, \alpha = 0.5, x = 0.005$, as in [23], [30].

$$C_{lead-lag} = K_c \left( \frac{s + 1/\lambda}{s + 1/(x\lambda)} \right)^\alpha = K_c x^\alpha \left( \frac{\lambda s + 1}{x\lambda s + 1} \right)^\alpha, 0 < x < 1 \tag{22}$$

The Bode magnitude and phase plots for the realizations of the FO lead-lag compensator at different orders are illustrated in Fig. 10. In this case too, an improvement in accuracy with increasing order up to 5 is observed. (Fig. 11 shows the CFE based fourth order realization accuracy at different fractional orders for FO lead-lag compensator. From Fig. 11 it can be seen that unlike the FO[PD] controllers, the FO lead lag compensators work well at higher frequencies as well. This is due to the drooping nature of the ideal response of the FO lead-lag compensator as opposed to a more flat response for the FO[PD] controllers at higher frequencies. Symbolic expression for 4th order realization of FO lead-lag compensator (22) is given by (23). Due to the high complexity of the symbolic expression for the 5th order realization, which is difficult to implement practically, its simulation example is excluded from the present work.

$$H_4 = \cfrac{K_c x^\alpha \left[\begin{array}{l} \left(\begin{array}{l} \alpha^4 x^4 - 50\alpha x^4 - 10\alpha^3 x^4 + 24x^4 + 35\alpha^2 x^4 + 384x^3 \\ +20\alpha^3 x^3 - 320\alpha x^3 + 40\alpha^2 x^3 - 4\alpha^4 x^3 - 150\alpha^2 x^2 \\ +6\alpha^4 x^2 + 864x^2 - 4\alpha^4 x + 40\alpha^2 x + 320\alpha x - 20\alpha^3 x \\ +384x + \alpha^4 + 24 + 10\alpha^3 + 50\alpha + 35\alpha^2 \end{array}\right)\lambda^4 s^4 \\[1em] +\left(\begin{array}{l} 480x^3 + 180\alpha^2 x^3 - 20\alpha^3 x^3 - 520\alpha x^3 + 60\alpha^3 x^2 + 2880x^2 \\ -960\alpha x^2 - 180\alpha^2 x^2 - 60\alpha^3 x - 180\alpha^2 x + 960\alpha x + 2880x \\ +520\alpha + 180\alpha^2 + 20\alpha^3 + 480 \end{array}\right)\lambda^3 s^3 \\[1em] +\left(\begin{array}{l} 180\alpha^2 x^2 - 1260\alpha x^2 + 2160x^2 - 360\alpha^2 x + 5760x \\ +180\alpha^2 + 1260\alpha + 2160 \end{array}\right)\lambda^2 s^2 \\[0.5em] +(-840\alpha x + 3360x + 840\alpha + 3360)\lambda s + 1680 \end{array}\right]}{\left[\begin{array}{l} \left(\begin{array}{l} \alpha^4 x^4 + 50\alpha x^4 + 10\alpha^3 x^4 + 24x^4 + 35\alpha^2 x^4 + 384x^3 \\ +40\alpha^2 x^3 + 320\alpha x^3 - 4\alpha^4 x^3 - 20\alpha^3 x^3 + 6\alpha^4 x^2 - 150\alpha^2 x^2 \\ +864x^2 - 320\alpha x + 384x + 20\alpha^3 x + 40\alpha^2 x - 4\alpha^4 x \\ +35\alpha^2 + \alpha^4 - 50\alpha - 10\alpha^3 + 24 \end{array}\right)\lambda^4 s^4 \\[1em] +\left(\begin{array}{l} 480x^3 + 180\alpha^2 x^3 + 20\alpha^3 x^3 + 520\alpha x^3 - 60\alpha^3 x^2 + 2880x^2 \\ +960\alpha x^2 - 180\alpha^2 x^2 + 60\alpha^3 x - 180\alpha^2 x - 960\alpha x + 2880x \\ -520\alpha + 180\alpha^2 - 20\alpha^3 + 480 \end{array}\right)\lambda^3 s^3 \\[1em] +\left(\begin{array}{l} 180\alpha^2 x^2 + 1260\alpha x^2 + 2160x^2 - 360\alpha^2 x + 5760x \\ +180\alpha^2 - 1260\alpha + 2160 \end{array}\right)\lambda^2 s^2 \\[0.5em] +(840\alpha x + 3360x - 840\alpha + 3360)\lambda s + 1680 \end{array}\right]} \tag{23}$$

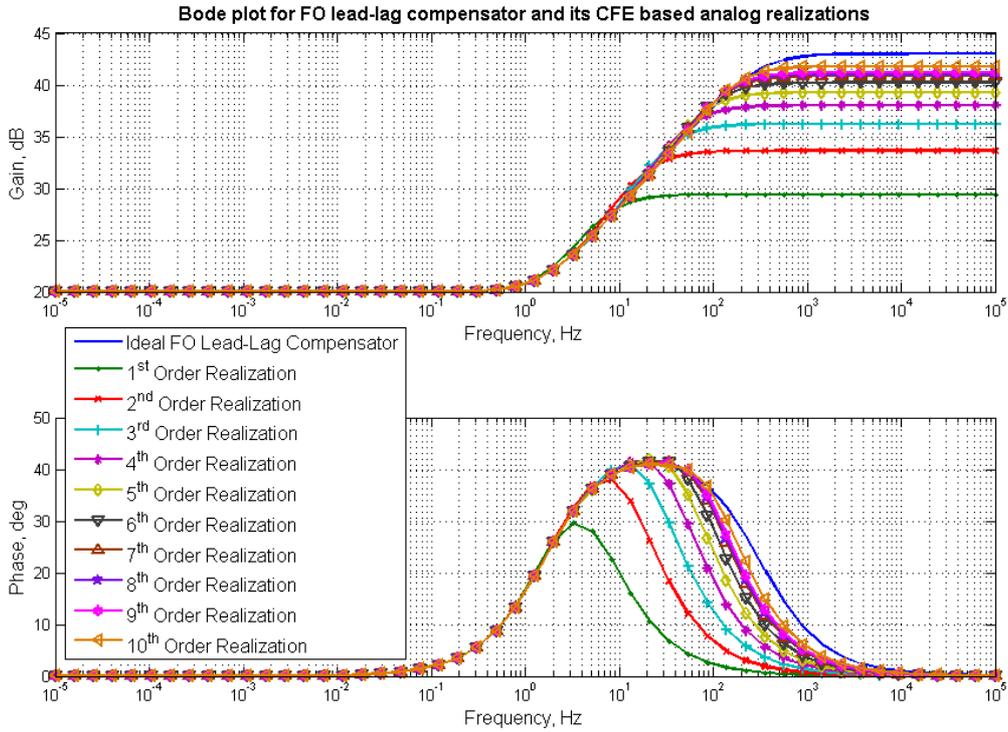

Fig. 10. CFE based analog approximation of Lead-Lag Compensator.

$$H_s = -\frac{\begin{aligned}&\left(\begin{matrix}\alpha^5 x^5 + 274\alpha x^5 + 85\alpha^3 x^5 - 120x^5 - 225\alpha^2 x^5 - 15\alpha^4 x^5 - 1005\alpha^2 x^4 - 5\alpha^3 x^4 - 3000x^4\\ -5\alpha^5 x^4 + 45\alpha^4 x^4 + 3250\alpha x^4 + 4000\alpha x^3 + 1230\alpha^2 x^3 - 30\alpha^4 x^3 - 12000x^3 + 10\alpha^5 x^3\\ -410\alpha^3 x^3 + 1230\alpha^2 x^2 - 4000\alpha x^2 - 10\alpha^5 x^2 - 30\alpha^4 x^2 - 12000x^2 + 410\alpha^3 x^2 - 3000x\\ -3250\alpha x + 45\alpha^4 x + 5\alpha^5 x - 1005\alpha^2 x + 5\alpha^3 x - 120 - 85\alpha^3 - 225\alpha^2 - \alpha^5 - 15\alpha^4 - 274\alpha\end{matrix}\right)\lambda^5 s^5\\ +\left(\begin{matrix}420\alpha^3 x^4 - 3600x^4 - 2130\alpha^2 x^4 - 30\alpha^4 x^4 + 4620\alpha x^4 - 36000x^3 + 120\alpha^4 x^3 - 1560\alpha^2 x^3\\ -840\alpha^3 x^3 + 21000\alpha x^3 - 180\alpha^4 x^2 + 7380\alpha^2 x^2 - 72000x^2 - 1560\alpha^2 x + 840\alpha^3 x\\ -36000x - 21000\alpha x + 120\alpha^4 x - 420\alpha^3 - 3600 - 4620\alpha - 2130\alpha^2 - 30\alpha^4\end{matrix}\right)\lambda^4 s^4\\ +\left(\begin{matrix}420\alpha^3 x^3 - 25200x^3 + 19740\alpha x^3 - 5040\alpha^2 x^3 + 5040\alpha^2 x^2 - 1260\alpha^3 x^2 - 126000x^2 + 31500\alpha x^2\\ +5040\alpha^2 x - 31500\alpha x - 126000x + 1260\alpha^3 x - 25200 - 420\alpha^3 - 5040\alpha^2 - 19740\alpha\end{matrix}\right)\lambda^3 s^3\\ +\left(\begin{matrix}30240\alpha x^2 - 3360\alpha^2 x^2 - 67200x^2 - 168000x + 6720\alpha^2 x - 30240\alpha\\ -3360\alpha^2 - 67200\end{matrix}\right)\lambda^2 s^2 + (-75600x + 15120\alpha x - 15120\alpha - 75600)\lambda s - 30240\end{aligned}}{\begin{aligned}&\left(\begin{matrix}\alpha^5 x^5 + 274\alpha x^5 + 15\alpha^4 x^5 + 120x^5 + 85\alpha^3 x^5 + 225\alpha^2 x^5 - 45\alpha^4 x^4 + 3250\alpha x^4 - 5\alpha^3 x^4 - 5\alpha^5 x^4\\ +3000x^4 + 1005\alpha^2 x^4 + 10\alpha^5 x^3 + 4000\alpha x^3 - 1230\alpha^2 x^3 + 12000x^3 - 410\alpha^3 x^3 + 30\alpha^4 x^3\\ -4000\alpha x^2 - 1230\alpha^2 x^2 - 10\alpha^5 x^2 + 410\alpha^3 x^2 + 30\alpha^4 x^2 + 12000x^2 + 5\alpha^3 x + 5\alpha^5 x\\ +3000x + 1005\alpha^2 x - 45\alpha^4 x - 3250\alpha x - 85\alpha^3 x + 225\alpha^2 - \alpha^5 - 274\alpha + 15\alpha^4 + 120\end{matrix}\right)\lambda^5 s^5\\ +\left(\begin{matrix}420\alpha^3 x^4 + 3600x^4 + 2130\alpha^2 x^4 + 30\alpha^4 x^4 + 4620\alpha x^4 + 36000x^3 - 120\alpha^4 x^3 + 1560\alpha^2 x^3\\ -840\alpha^3 x^3 + 21000\alpha x^3 + 180\alpha^4 x^2 - 7380\alpha^2 x^2 + 72000x^2 + 1560\alpha^2 x + 840\alpha^3 x\\ +36000x - 21000\alpha x - 120\alpha^4 x - 420\alpha^3 + 3600 - 4620\alpha + 2130\alpha^2 + 30\alpha^4\end{matrix}\right)\lambda^4 s^4\\ +\left(\begin{matrix}420\alpha^3 x^3 + 25200x^3 + 19740\alpha x^3 + 5040\alpha^2 x^3 - 5040\alpha^2 x^2 - 1260\alpha^3 x^2 + 126000x^2 + 31500\alpha x^2\\ -5040\alpha^2 x - 31500\alpha x + 126000x + 1260\alpha^3 x + 25200 - 420\alpha^3 + 5040\alpha^2 - 19740\alpha\end{matrix}\right)\lambda^3 s^3\\ +(30240\alpha x^2 + 3360\alpha^2 x^2 + 67200x^2 + 168000x - 6720\alpha^2 x - 30240\alpha + 3360\alpha^2 + 67200)\lambda^2 s^2\\ +(75600x + 15120\alpha x - 15120\alpha + 75600)\lambda s + 30240\end{aligned}} K_c x^\alpha$$

$$\tag{24}$$

The convergent polynomial of 3$^{\text{rd}}$ order obtained using CFE is as follows

$$Q(s) = \frac{64.88s^3 + 798.69s^2 + 2478.46s + 2203.66}{s^3 + 34.68s^2 + 177.63s + 220.36} \tag{25}$$

This on expanding by CFE again will lead to the domino ladder form as follows

$$C(s) = 64.88 + \cfrac{1}{-6.88 \times 10^{-4}s - 0.01 + \cfrac{1}{181.35s + 752.74 + \cfrac{1}{-0.01s - 0.03}}} \tag{26}$$

Comparing with the domino ladder form, we have:

$$Z_1(s) = 64.8883$$

$$Y_2(s) = -6.8879 \times 10^{-4}s - 0.0196$$

$$Z_3(s) = 181.3516s + 752.7414$$

$$Y_4(s) = -0.0179s - 0.0372$$

Even though the admittance values obtained do not pertain to any standard circuit element, they can be realized by cascading an all pole filter with a phase lead circuit. For example, let us take the admittance of the domino ladder circuit as

$$Y(s) = A + Bs, \quad \Rightarrow Z(s) = \frac{1}{A + Bs} \tag{27}$$

Substituting $s = j\omega$ in the above equation

$$Z(j\omega) = \frac{1}{A + j\omega B} \tag{28}$$

By simple algebraic manipulation we obtain

$$Z(s) = \frac{1}{A + Bs} = \frac{1}{A - Bs} \times \frac{A - Bs}{A + Bs} = Z_1(s) \times Z_2(s) \tag{29}$$

Thus $Z(s)$ can be obtained by cascading two separate transfer function blocks $Z_1(s)$ and $Z_2(s)$. $Z_1(s)$ is a first order filter with unstable pole, while $Z_2(s)$ can be realized by a lag-lead circuit with non-minimum phase zero.

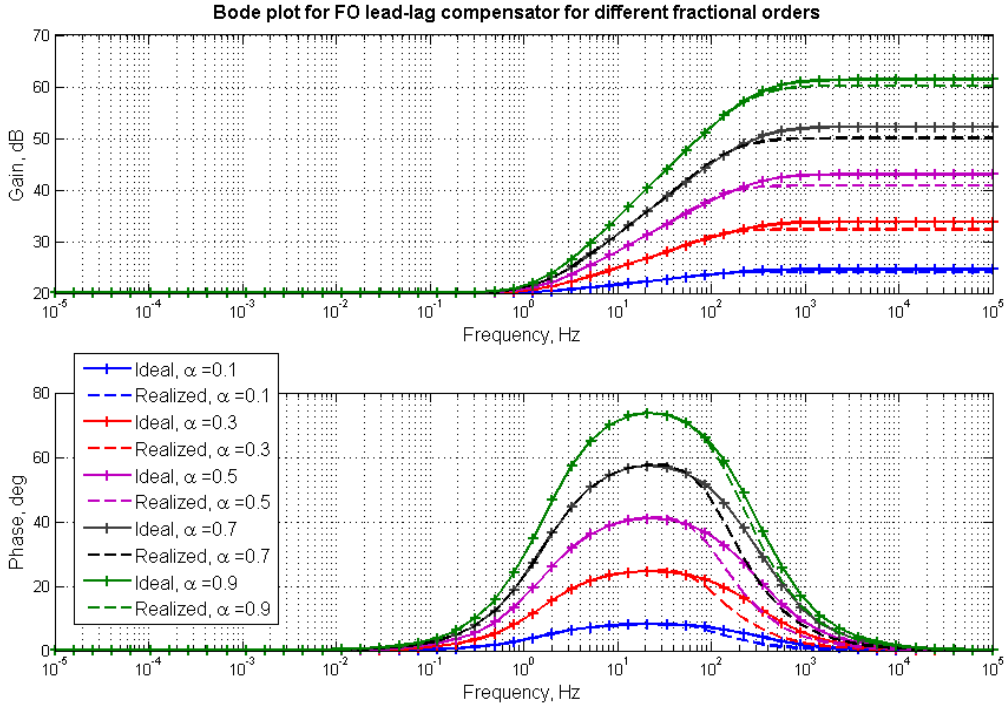

Fig. 11. Frequency domain fitting at different α for FO lead-lag compensator.

### 3.5. Discussion and the contributions

The main contribution of the paper is to give the symbolic expressions of the rationalized transfer functions of the FO controllers, parameterized in terms of the controller parameters. The main advantage of the proposed method is that with the knowledge of the controller gains and orders, one can get the higher order rationalized transfer functions without running CFE. In addition, for the first time we report the analog realization of fractional powers of rational transfer functions $[G(s)]^{\pm\lambda}$. Although in literature the rational approximation of basic FO element ($s^{\pm\lambda}$) is adequately addressed, but to the best of our knowledge there is no evidence of rational approximation of FO lead lag compensator and FO[PD] to facilitate their circuit realization. Although Carlson's method can be used to rationalize these type of transfer function as previously studied in [17] but it suffers from low accuracy. Two balance between these two issues – realizability and accuracy, especially for the fractional powers of rational transfer functions, the CFE scores better over the other variants. The aim of the present is not to realize the already tuned FO controllers in terms of standard circuit elements. It is to be note that if the accuracy of the realization is not good around the gain and phase cross over frequencies it can affect the gain and phase margins of the system under control.

## 4. Conclusion

CFE based analog realization of FO[PD], FOPID controller and FO lead-lag compensator have been reported in the present work. Increasing order of approximation leads to a better accuracy but higher complexity of the models. Hence a trade-off between accuracy and practical realizability is adopted. Domino ladder based circuit realization techniques using practical impedance values for these realizations are attempted and some illustrative examples have been shown to obtain these impedances. The work is of practical significance to circuit designers who would be interested in the hardware implementation of popular fractional order controller structures. It shows the various intricacies involved in the circuit designing process and the corresponding trade-off between the order of accuracy and the complicacy of the realized filter. The symbolic transfer functions, represented in terms of the free controller tuning parameters, help to directly obtain the analog approximations and the corresponding impedance values via CFE. In future, comparison of analog and digital realization of these FO controller structures and their complexity can be studied.